\documentclass[aps,prl,floatfix,superscriptaddress,twocolumn,footinbib]{revtex4-1}
\usepackage{amssymb}
\usepackage{amsmath}
\usepackage{amsfonts}
\usepackage{appendix}
\usepackage{bm}
\usepackage{graphicx}
\usepackage{epsfig}
\usepackage{epstopdf}
\usepackage{balance}
\usepackage[dvipsnames]{xcolor}
\usepackage{calc}
\usepackage{natbib}
\usepackage[colorlinks,
            linkcolor=blue,
            anchorcolor=blue,
            citecolor=blue,
            urlcolor=blue]{hyperref}
\usepackage{lipsum}

\begin{document}
\title{Topological flat bands in twisted trilayer graphene}
\author{Zhen Ma}
\affiliation{School of Physics and Wuhan National High Magnetic Field Center,
Huazhong University of Science and Technology, Wuhan 430074,  China}
\author{Shuai Li}
\affiliation{School of Physics and Wuhan National High Magnetic Field Center,
Huazhong University of Science and Technology, Wuhan 430074,  China}
\author{Ya-Wen Zheng}
\affiliation{School of Physics and Wuhan National High Magnetic Field Center,
Huazhong University of Science and Technology, Wuhan 430074,  China}
\author{Meng-Meng Xiao}
\affiliation{School of Physics and Wuhan National High Magnetic Field Center,
Huazhong University of Science and Technology, Wuhan 430074,  China}
\author{Hua Jiang}
\affiliation{School of Physical Science and Technology, Soochow University, Suzhou 215006, China}
\author{Jin-Hua Gao}
\email{jinhua@hust.edu.cn}
\affiliation{School of Physics and Wuhan National High Magnetic Field Center,
Huazhong University of Science and Technology, Wuhan 430074,  China}
\author{X. C. Xie}
\affiliation{International Center for Quantum Materials, School of Physics, Peking University, Beijing 100871, China }
\affiliation{Collaborative Innovation Center of Quantum Matter, Beijing 100871, China}
\affiliation{CAS Center for Excellence in Topological Quantum Computation, University of Chinese Academy of Sciences, Beijing 100190, China}
\begin{abstract}
Twisted trilayer graphene (TLG) may be the simplest realistic system so far, which has flat bands with nontrivial topology.  
Here, we give a detailed calculation about its band structures and the band topology, i.e. valley Chern number of the nearly flat bands,  with the continuum model. With realistic parameters, the magic angle of twisted TLG is about  1.12$^{\circ}$, at which two nearly flat bands appears. Unlike the twisted bilayer graphene, a small twist angle can induce a tiny gap at all the Dirac points, which can be enlarged further by a perpendicular electric field. The  valley Chern numbers of the two nearly flat bands in the twisted TLG  depends on the twist angle $\theta$ and the perpendicular electric field $E_\perp$.   Considering its topological flat bands,  the twisted TLG should be an ideal experimental platform to study the strongly correlated physics in topologically nontrivial flat band systems.  And, due to its reduced symmetry, the correlated states in twisted TLG should be quite different from that in twisted bilayer graphene and twisted double bilayer graphene. 
\end{abstract}
\maketitle

\emph{Introduction.}---Flat bands with  nontrivial topology is believed to be the key in realizing high-temperature fractional topological states in the absence of the magnetic field\cite{wen2011,sun2011,mudry2011,wangyifei2011}. Due to the quenched kinetic energy of the flat bands, the interplay between the Coulomb interaction and nontrivial band topology can induce novel topological strongly correlated electronic states\cite{liu2013,Neupert2015}. However, most of the studies so  far are based on theoretical flat band lattice models. To find topological flat bands in  real materials is still a big challenge for both theory and experiment. 

In last few years, the flat bands in twisted bilayer graphene (BLG) have drawn lots of research interest. 
 In twisted BLG, the twisting produces a long-period moir\'e pattern, and induce moir\'e bloch bands\cite{prl2007}.  Importantly, near some magic angles, the bands near the Fermi level become nearly flat, and thus  the Coulomb interaction begin to play a dominating role\cite{mac2011}. Recently,  correlated insulating phase and  superconducting phase in twisted BLG  have been observed in experiment\cite{cao2018a,cao2018b,science2019}, and a lot of efforts have been made to theoretically understand the correlated states in twisted BLG\cite{Po2018,fuliang2018,fuliang20182,xu2018,Isobe2018,kang2018,wu2018,Mac2018,weiqiang2018}.  

Very recently, several theoretical works point out the topological flat bands can be realized in the twisted chiral graphene multilayer systems, where two chiral stacked graphene multilayers are placed on top of each other with a small twisting angle\cite{yhzhang2019,koshino2019,ashvin2019,liu20192,jeil2019}.  The most studied example is  the twisted double bilayer graphene (AB/AB and AB/BA),  in which  twisting can induce two nearly flat bands with  nonzero valley Chern number. The Chern number can be further controlled by a perpendicular electric field.  Interestingly,  the twisted double bilayer graphene (BLG) has already been realized in experiment, and superconductivity and correlated insulating states are discovered near  some magic angles\cite{kim2019}. Note that twisting induced topological flat bands are also predicted to exist in the trilayer graphene (TLG)/Boron-Nitride heterostructure\cite{wangfeng2019,bnprl2019}. 

\begin{figure*}
\centering
\includegraphics[width=17.5cm]{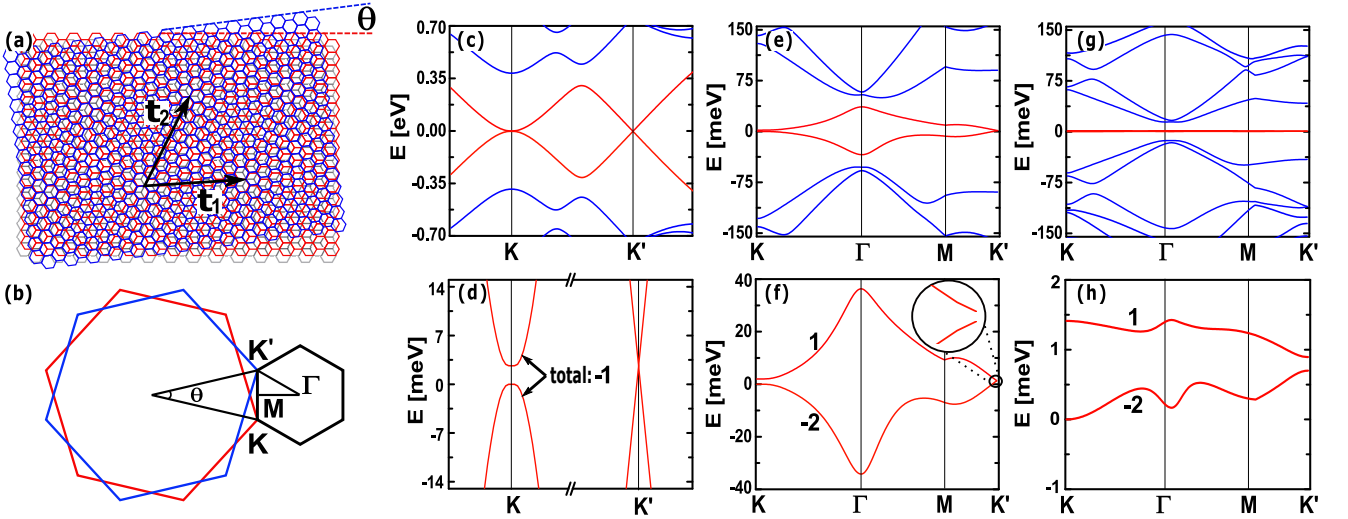}
\caption{Band structure of the twisted trilayer graphene with the minimal model. (a) Lattice structure of the twisted trilayer graphene.  $\theta$ is the twisting angle between the top monolayer (blue, n=3) and a AB-stacked bilayer(n=1,2). $t_1$ and $t_2$ is the moir\'e pattern lattice vertor. n is layer number. (b) Brillouin zone. Blue lines denote the BZ of the top monolayer, red lines denote the BZ of the bilayer, and the black lines represent the moir\'e BZ of the twisted trilayer graphene. (c) and (d)
The band structure at $\theta=5^{\circ}$. (e) and (f) The band structure at $\theta=1.53^\circ$. (g) and (h) The band structure at $\theta=1.12^\circ$. The parameters of the minimal model are: $\gamma_0$=2.464 eV, $\gamma_1$=400 meV, $\gamma_3$=$\gamma_4$=$\Delta'$=0. The black numbers in (d), (f), (h) are the valley Chern numbers  for  two central bands (red lines).}
\label{fig1}
\end{figure*}

In this work, we theoretically study a simpler topological flat band system, i.e.,\emph{the twisted TLG}, where one graphene monolayer is stacked on the top of a graphene bilayer with a small twisting angle. The twisted TLG  also has two topological nontrivial flat bands with electric-field controlled valley Chern number near the magnetic angle. 
We numerically calculate the band structures of the twisted TLG based on the continuum model\cite{tlg2013}. Our main findings are:  (a) Like the twisted BLG, twisted TLG also has some magic angles, near which the twisting can produce two nearly flat bands. The largest magnetic angle is about $1.12^\circ$. (b) Unexpectedly, even with a graphene monolayer as a building block, twisting can open a small gap at all the Dirac points  of the twisted TLG near the magic angle, including the Dirac points from the monolayer.  
Thus, near the magnetic angle, we get two separated nearly flat bands around the Fermi level.
The twisting opened gap here depends the twisting angle $\theta$, and  can be  enlarged by a perpendicular electric field $E_\perp$.  This is quite unlike  the twisted BLG, in which a perpendicular electric field can not open a gap at the Dirac points belong to the graphene monolayer, due to the $C_2T$ symmetry.  (c)  Further calculations show that such two flat bands in the twisted TLG are flat Chern bands with finite valley Chern number, even in the absence of the perpendicular electric field.   The valley Chern number  of the flat bands here are sensitive to the relative small tight-binding parameters of the graphite, e.g.  the triangle warping. Meanwhile,  the flat band Chern number  can be controlled by the  $\theta$ and $E_{\perp}$.  
 Finally, because of the reduced symmetry of the  twisted TLG resulted from its asymmetry stacking,   we thus expect that  the twisted TLG should have different topological correlated states from that  in twisted  BLG and double BLG.

\emph{Lattice structure.}---As illustrated in Fig.~\ref{fig1} (a),  we consider a twisted trilayer graphene, where one graphene monolayer (blue) is stacked on top of a AB-stacked bilayer graphene (red) with a small twist angle $\theta$.  At $\theta=0$, we get a AAB-stacked trilayer graphene, where   $\bm{a}_1=a(1/2,\sqrt{3}/2)$ and $\bm{a}_2=a(-1/2,\sqrt{3}/2)$ are the lattice vectors in the non-rotated case, and the corresponding reciprocal lattice vector are
$\bm{b}_1=\frac{4\pi}{3|\bm{a}_1|}(2\bm{a}_1-\bm{a}_2)$ and $\bm{b}_2=\frac{4\pi}{3|\bm{a}_1|}(-\bm{a}_1+2\bm{a}_2)$ . 
$a\approx0.246$ nm is the lattice constant of graphene.  Similar as the case of twisted BLG, a commensurate moir\'e supercell can be formed under the condition of $\cos\theta (m)=(3m^2+3m+1/2)/(3m^2+3m+1)$. Here, $m$ is a positive integer. $\bm{t}_1=m\bm{a}_1+(m+1)\bm{a}_2$ and $\bm{t}_2=-(m+1)\bm{a}_1+(2m+1)\bm{a}_2$ are the lattice vectors of the moir\'e supercell [see Fig.~\ref{fig1} (a)]. The corresponding reciprocal lattice vector of the moir\'e pattern are $\textbf{G}_1=\frac{4\pi}{3|\bm{t}_1|}(2\bm{t}_1-\bm{t}_2)$ and $\textbf{G}_2=\frac{4\pi}{3|\bm{t}_1|}(-\bm{t}_1+2\bm{t}_2)$. The Dirac points are $K^{l}_{\xi}=-\xi R(\theta_l)(2\bm{b}_1+\bm{b}_2)/3$, where $l=t$ ($b$) denotes the Dirac point belongs to the top monolayer (bottom BLG). $R(\theta)$ is the rotation matrix, and $\theta_{t}=\theta/2$ ($\theta_{b}=-\theta/2$). $\xi=\pm{1}$ is the valley index.   For convenience, we define $K=K^{t}_{+}$ and $K'=K^{b}_{+}$, as illustrated in Fig. \ref{fig1} (b).

\textit{Continuum Hamiltonian.}---To calculate the electronic band structure of the twisted TLG, we use the method in Ref. \onlinecite{mac2011}, which was generalized to study the twisted multilayer graphene\cite{yhzhang2019,koshino2019,ashvin2019,liu20192,jeil2019}. With the Bloch basis $(A_1,B_1, A_2,B_2,A_3, B_3)$, the continuum Hamiltonian of the twisted TLG is a $6\times6$ matrix. For example, $A_3$ is the Bloch function of the carbon $p_z$ orbital on the sublattice A in the third layer [top blue layer in Fig. \ref{fig1} (a)], i.e., $|\bm{k},A_3\rangle=N^{-1/2}\sum_{\bm{R}_i \in A_3 }e^{i\bm{k}\cdot \bm{R}_{i}} |\phi_{p_z}(\bm{R}_i)\rangle$. The continuum Hamiltonian of one valley is 
\begin{equation}\label{H1}
H_{\textrm{tTLG}}(\theta)=\left( \begin{array}{ccc}
 h_b(k_1)& T(\bm{r}) \\
   T^\dagger(\bm{r})&  h_0(k_2)
\end{array} \right)+U
\end{equation} 
where $k_1=R({-\theta/2})(k-K_{\xi}^{b})$ and $k_2=R({\theta/2})(k-K_{\xi}^{t})$.
$h_0(\bm{k})=-\hbar v_F \bm{k} \cdot \bm{\sigma}$  is the Hamiltonian of monolayer graphene, $h_b(k)$ is the Hamiltonian of AB-stacked BLG,  the off-diagonal term $T$ represents the coupling between the twisted graphene monolayer  and bilayer, and $U=diag(-V, -V, 0, 0, V, V )$ is the perpendicular electric field induced asymmetry potential between layers. 

For the AB-stacked BLG\cite{bilayer2013}, 
\begin{equation}\label{hb}
h_{b}(\bm{k})=\left( \begin{array}{ccc}
 h_0(k)& g(k)^\dagger  \\
   g(k)&  h_0(k)
\end{array} \right)+h_{\Delta'}.
\end{equation} 
Here, $g(k)$ represent the interlayer hopping in BLG,  
\begin{equation}\label{gk}
g(k)=\left( \begin{array}{cc}
\hbar v_4k_+ &\gamma_1\\
\hbar v_3k_- &\hbar v_4k_+ 
\end{array} \right),
\end{equation} 
where $k_\pm=\xi k_x\pm ik_y$, $v_{3,4}=\sqrt{3}a\gamma_{3,4}/2\hbar$.  $\gamma_1$ is the vertical hopping. $\gamma_3$ and $\gamma_4$ are smaller remote hopping, which corresponds to trigonal warping and electron-hole asymmetry, respectively. We also include the energy difference between dimmer and non-dimmer sites in BLG\cite{bilayer2013}, i.e., $h_{\Delta'}=diag(0,\Delta',\Delta',0)$.

The mori\'e interlayer coupling  is $T(\bm{r})=\sum_{n=0,1,2} T_n\cdot e^{-i\bm{Q}_n\cdot \bm{r}}$, where
\begin{equation}\label{T}
T_n=\left( \begin{array}{cc}
 0 & 1  \\
\end{array} \right)\otimes\left( \begin{array}{cc}
 \omega_1 & \omega_2e^{in\phi} \\
 \omega_2e^{-in\phi} & \omega_1
\end{array} \right).
\end{equation} 
Here, $\phi=2\pi/3$, $\omega_1$ and $\omega_2$ are the intrasublattice and intersublattice tunneling amplititude between the adjacent twisted layers, where  $\omega_1 < \omega_2$ due to the atomic corrugations\cite{prx2018}.
Note that, the hopping matrix $T_n$ couples the Bloch states from adjacent twisted graphene layers  with the momentum difference $\bm{Q}_n=R(n\phi)\cdot (\bm{K}-\bm{K'})$.

\textit{Topological flat bands with the minimal model.}---Let us first consider a minimal model of the twisted TLG, where the smaller parameters of the BLG, i.e. $\gamma_{3,4}$ and $\Delta'$ in Eq.\eqref{hb}, are ignored. As we will see, in twisted TLG, the topology of the twist induced flat bands (i.e the valley Chern number) are sensitive to the choice of these small parameters. Thus, the minimal model here is a good starting point.

The calculated band structures of the twisted TLG with different twisted angle $\theta$ are given in Fig. \ref{fig1} (c-h).  Generally speaking, the electronic structure of the twisted TLG near the Dirac points can be viewed as a combination of a AB-stacked BLG and a graphene monolayer\cite{tlg2013} [see also in Fig. \ref{fig1} (c)]. For one valley in the mori\'e Brillouin zone, there are two inequivalent Dirac points [$K$ and $K'$ in Fig. \ref{fig1} (b)]. The electronic states near $K'$ is mainly located on the twisted graphene monolayer, where two linear bands form a Diarc cone like graphene. Meanwhile, there are two parabolic bands near the $K$ points like the BLG, the electronic states in which are distributed on the BLG part of the twisted TLG.  Similar as the twisted BLG, as the twisted angle is reduced, the two bands near the Fermi level narrows [see Fig. \ref{fig1} (d)]. At a magic angle about $1.12^\circ$, these two bands become nearly flat[see Fig. \ref{fig1} (d)]. Note that twisting can open a small gap between the two  parabolic bands near the $K$ point [see in Fig. \ref{fig1} (d)]. The smaller $\theta$ is, the larger the gap is.   These band features above are consist with the former understanding about the twisted TLG\cite{tlg2013}.

Our calculations show some unusual characteristics of the band structure of twisted TLG. First, unlike the twisted BLG, the Dirac cone at $K'$ from the top single layer is gaped by a twisting angle. Though this gap is tiny, it becomes distinguishable when $\theta$ approaches the magic angle, as shown in Fig. \ref{fig1} (f), (h). And the gap can be further enlarged by a perpendicular electric field $E_\perp$. So, the two bands near the Fermi level in the twisted TLG are actually separated by the twisting, even in the absence of the perpendicular electric field. This is quite different from the case of twisted BLG, where the linear dispersion near the Dirac points of single layer graphene is always retained, and an applied $E_{\perp}$ only induce a potential difference between layers. Note that, in twisted BLG, the degeneracy at Dirac points are protected by the $C_2T$ symmetry, which is lack in twisted TLG. 

\begin{figure}[t!]
\centering
\includegraphics[width=8.5cm]{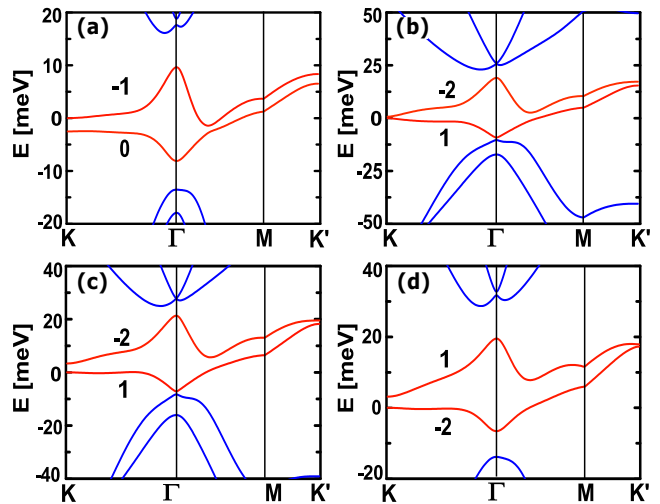}
\caption{Band structure of twisted twisted trilayer graphene with the additional band parameters. Here, $\theta=1.12^\circ$. The additional band parameters are: trigonal warping $\gamma_3$, particle-hole asymmetry $\gamma_4$ and on-site potential of dimer sites $\Delta'$. (a) Influence of the trigonal warping $\gamma_3$. The parameters are  $\gamma_0/\gamma_1/\gamma_3/\gamma_4/\Delta'=2464/400/200/0/0$ meV. (b) Influence of the particle-hole asymmetry $\gamma_4$. The parameters are $\gamma_0/\gamma_1/\gamma_3/\gamma_4/\Delta' =2464/400/320/138/0$ meV. (c) and (d) The band structures of twisted trilayer graphene with two set of parameters commonly used in literatures.  
The parameters in (c): $\gamma_0/\gamma_1/\gamma_3/\gamma_4/ \Delta'=2464/400/320/44/50 $ meV\cite{McCann_2013}. The parameters in (d): $\gamma_0/\gamma_1/\gamma_3/\gamma_4/ \Delta'=2610/360/283/138/15$ meV\cite{Mac2014}. Black numbers are the valley Chern number.}
\label{fig2}
\end{figure}

\begin{figure*}[t]
\centering
\includegraphics[width=16cm]{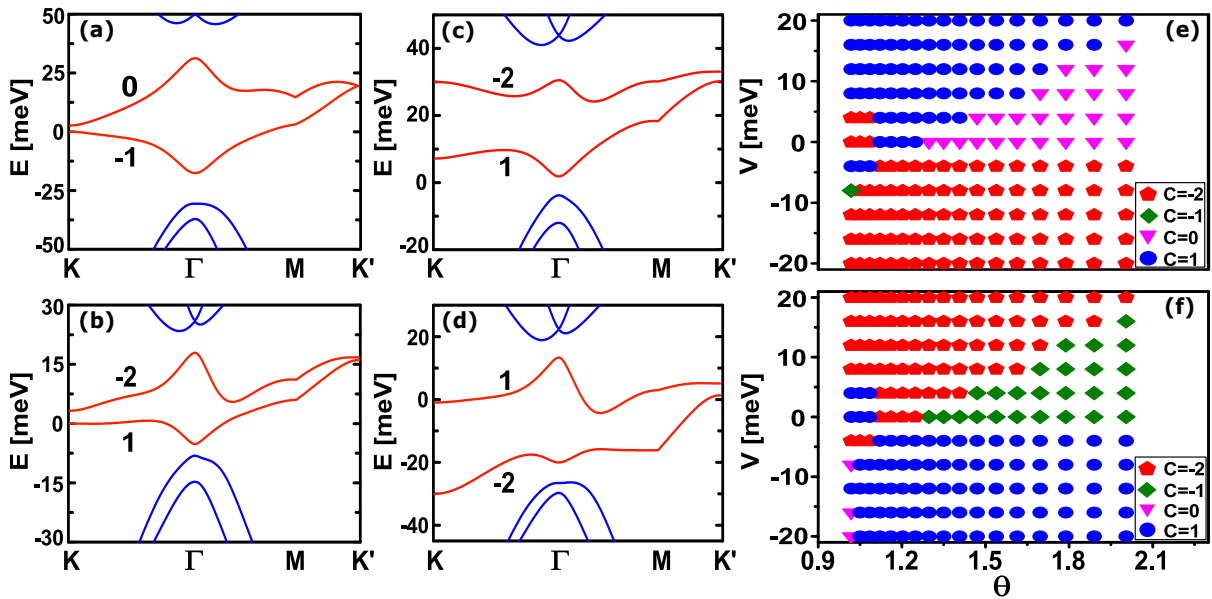}
\caption{ (a) and (b) Band structures of the twisted trilayer graphene at $\theta=1.33^\circ$ and $\theta=1.05^\circ$, respectively. (c) and (d) Band structure of the twisted trilayer graphene at $\theta=1.12^\circ$ with $V=20$ meV and $V=-20$ meV, respectively. 
(e) and (f) Chern number of the twisted trilayer graphen for electron band and hole band, respectively. The parameters: $\gamma_0/\gamma_1/\gamma_3/\gamma_4/ \Delta'=2610/360/283/138/15$ meV.}
\label{fig3}
\end{figure*}

The two flat bands in twisted TLG have nonzero valley Chern number. Due to the gaps at the Dirac points, we can numerically calculate the valley Chern number of each flat band $C_{n\xi}$, where $n=e,h$ is the band index (electron band or hole band) and $\xi=\pm$ denotes the valley. We use the standard formula, where the Berry curvature is
 \begin{equation}
     \Omega_n(\vec{k})=-2\sum_{n'\neq n'}
\textrm{Im}[\frac{\langle u_n|\frac{\partial H}{\partial k_x}|u_{n'}\rangle\langle u_{n'}|\frac{\partial H}{\partial k_y}|u_{n}\rangle}{(E_{n'}-E_{n})^2}].
 \end{equation}
  $|u_{n}\rangle$ is the moir\'e superlattice Bloch state, and  $E_n$ is the corresponding  eigenvalue.  The Chern number of the $n$th band $C_n$  is calculated through $C_n=\int_{\textrm{mBZ}}d^2\vec{k}\Omega_n(\vec{k})/2\pi$. 
  In  Fig. \ref{fig1} (f) and (h), the black numbers are the calculated Chern numbers of the electron band ($C_{e+}=1$) and hole band ($C_{h+}=-2$) in one valley ($\xi=+1$) with the minimal model. Note that, due to the time reversal symmetry, $C_{n+}=-C_{n-}$. When $\theta$ is large, e.g. $\theta=5^\circ$ in Fig.~\ref{fig1} (c) and (d),  the gap at $K'$ is too tiny to be distinguished. We thus calculate the total Chern number of the two low energy flat bands. We find that $C_{\textrm{total}}=-1$, as shown in Fig. \ref{fig1} (d). Note that, $C_{\textrm{total}}=-1$ is always valid in the twisted TLG. Actually, there is a  rule of the total Chern number in twisted chiral multilayer graphene, which is reported in a separated work by authors and collaborators\cite{liu20192}.

 \emph{Influence of additional band parameters.}--- Since the bandwidth of the two central bands is narrow at small twist angle, the additional bands parameter of BLG, i.e. $\gamma_3$, $\gamma_4$ and $\Delta'$, can essentially influence the behaviors of the two flat bands, thus change the band topology as well.  So, when we consider the  topology of the flat bands, the choice of these additional band parameters can give rise to different conclusions. 
 
 In Fig.~\ref{fig2}, we show the influence of the additional band parameters on the flat band topology. We first consider the trigonal warping $\gamma_3$ in Fig. \ref{fig2} (a). For the Chern number of the two central bands ($C_{e+}$, $C_{h+}$), there is a change from (1,-2) to (-1,0), when we increase $\gamma_3$ from 0 to its realistic value  about 320 meV.  This is because that  a band touching between the two central bands occurs when increasing the $\gamma_3$. $\gamma_4$ and $\Delta'$ can affect the band topology in the similar way. The change of Chern number resulted from $\gamma_4$ is illustrated in Fig. \ref{fig2} (b). Finally, we use two sets of tight-binding parameters of the BLG, which are commonly used in literatures\cite{McCann_2013,Mac2014}, to calculate the band structure and valley Chern number of the twisted TLG [see Fig.~\ref{fig2} (c), (d)]. We see that a deviation of $\gamma_4$ less than 100 meV  can lead to different predictions about the valley Chern number.  In the following, we use the parameters in Fig. \ref{fig2} (f).

\emph{Phase diagram of Chern number.}---Here, with realistic parameters, we calculate the valley Chern number of the twisted TLG as a function of $\theta$ and $V$. The calculated results are given in Fig.~\ref{fig3}. Importantly, the twisted angle $\theta$ and the perpendicular electric field $E_\perp$ (i.e., V) are all  tunable in experiment, so that this phase diagram can be verified in further experiments. 

In Fig.~\ref{fig3} (a) and (b), we change $\theta$ from $1.61^\circ$ to $1.05^\circ$. As expected, twisting  not only can modify the Chern number of each central bands, but also change the gap at the Dirac points. We further calculate the band structures with different applied potential $V$.   Fig. \ref{fig2} (c) and (d) show the bands with $V=\pm$ 20 meV, respectively. Note that, the twisted TLG is a asymmetrical stacking, i.e. a single layer on a bilayer. So, the effects of $E_\perp$ on the band structure depend on its direction, i.e. $\pm V$ give rise to different band structures. This is different from the twisted double BLG, in which $E_\perp$ with opposite direction will give the same band structure, due to its symmetrical stacking.  Take the case of Fig. \ref{fig3} (c) for example, the  bandwidth of upper band is about $8.8$ meV, and that of the lower band is about $30$ meV. These are the typical bandwidth of the central bands in twisted TLG. With a finite $V$, the  electron flat band becomes
much narrower than the hole flat band, because of the electron-hole asymmetry\cite{koshino2019}. 
 Meanwhile, we see that the gap at $K'$ is enlarged to 3.8 meV [Fig. \ref{fig3} (d)] by an applied potential difference $V=20$ meV, while its typical value is about 0.2 meV when $V=0$ [Fig. \ref{fig1} (h)]. 

Finally, we give the phase diagram of the valley Chern number for each central band in Fig. \ref{fig3} (e) and (f). The calculated results illustrate that nearly flat bands (electron or hole bands) with Chern number from -2 to 2 can be realized in the twisted TLG system with proper twisting angle and perpendicular electric field.   

\emph{Summary.}---We have theoretically shown that twisted TLG has two separated flat bands with finite valley Chern number, which can be controlled by the twisting angle and applied potential difference. Considering its simpler structure and different symmetry, we think that the twist TLG is also a promising platform to study the novel correlated states in topological bands, which should be of the equal importance as the twisted double BLG.  We hope that our work can stimulate further essential experiment progress on this novel system.

\begin{acknowledgments}
We thank the supports by the National Natural Science Foundation of China (Grants No. 11534001, 11874160, 11274129, 11874026， 61405067), and the Fundamental Research Funds for the Central Universities (HUST: 2017KFYXJJ027), and NBRPC (Grants No. 2015CB921102). We thank Jianpeng Liu and Jia-Qi Cai for invaluable discussions. 
\end{acknowledgments}

\bibliography{twistedgraphene}
\end{document}